\documentclass[a4paper,reqno,superscriptaddress,12pt]{revtex4}
\usepackage[centertags]{amsmath}
\usepackage{amsfonts}
\usepackage{amssymb}
\usepackage{amsthm}
\usepackage{newlfont}
\usepackage{stmaryrd}
\usepackage{mathrsfs}
\usepackage{mathtools}
\usepackage{euscript}
\usepackage{graphicx}
\usepackage{enumerate}
\usepackage{todonotes}
\usepackage{comment}
\usepackage{graphicx}
\usepackage{epsfig}

\usepackage[utf8]{inputenc}

\newcommand{\opunit}{\text{1}\kern-0.22em\text{l}}

\newcommand{\dx}{\text{d}x}

\DeclareMathAlphabet{\mathpzc}{OT1}{pzc}{m}{it}

\newcommand{\id}{\textrm{d}}

\def\bea{\begin{eqnarray}}
\def\eea{\end{eqnarray}}
\def\ba{\begin{array}}
\def\ea{\end{array}}

\usepackage{changes}

\begin{document}

\title{Active processes in one dimension}

\author{Thibaut Demaerel and Christian Maes\\
{\it Instituut voor Theoretische Fysica, KU Leuven}}


\begin{abstract} We consider the thermal and athermal overdamped motion of particles in 1D geometries where discrete internal degrees of freedom (spin) are coupled with the translational motion.  Adding a driving velocity that depends on the time-dependent spin constitutes the simplest model of active particles (run-and-tumble processes) where the violation of the equipartition principle and of the Sutherland-Einstein relation can be studied in detail even when there is generalized reversibility. We give an example (with four spin values) where the irreversibility of the translational motion manifests itself only in higher-order (than two) time correlations. We derive a generalized telegraph equation as the Smoluchowski equation for the spatial density for an arbitrary number of spin values.  We also investigate the Arrhenius exponential law for run-and-tumble particles; due to their activity the slope of the potential becomes important in contrast to the passive diffusion case and activity enhances the escape from a potential well (if that slope is high enough). Finally, in the absence of a driving velocity, the presence of internal currents such as in the chemistry of molecular motors may be transmitted to the translational motion and the internal activity is crucial for the direction of the emerging spatial current.
\end{abstract}

\date{\today}
\maketitle


\section{Introduction}
With some abstraction, active particles may be considered objects whose spatial motion is coupled to internal degrees of freedom.  We refer to \cite{mar,be} for general reviews. Typical features of the motion include its persistence in direction.  Well-known examples are Janus particles or  self-propelled micro- and nanomotors where the internal degrees of freedom are coupled to a nonequilibrium environment, \cite{pdb}.  The orientation giving ``direction'' to the particle motion is governed by internal rotational diffusion (such as for active Brownians) or by a Poisson process (such as for run-and-tumble processes) over different chemo-mechanical states.\\
A different perspective is to think of adding other than just thermal noise, having its origin in the dynamics of other degrees of freedom. A run-and-tumble process in one dimension (see \cite{ind} for recent work) is thus viewed as a case of dichotomous noise, where one imagines a particle carrying a spin undergoing a flip dynamics, \cite{Kac,cv1,cv2}.  At fixed spin value $\sigma$ the particle continues in the same direction until the spin flips (at Poisson times).  On the lattice that makes a persistent random walk.  The Master equation of a persistent random walk, which can be seen as a walker on two lanes, is found in \cite{gold}. In the continuum limit the telegraph equation appears, for a density $u=u(x,t)$ on the line,
\begin{equation}\label{te}
\frac{\partial^2 u}{\partial t^2} + \eta\, \frac{\partial u}{\partial t}= c^2\,\frac{\partial^2 u}{\partial x^2}, \qquad \eta>0,
\end{equation}
and was first applied by Kelvin for calculating the impedance to be added to a cable to ensure transmission of a signal without changing its shape.  Indeed, for $\eta=0$ the equation is hyperbolic and can be used for ballistic transport.  On the other hand, taking into account dissipation we have $\eta>0$ and while keeping $c^2/\eta$ constant, the limit  $c^2,\eta\uparrow \infty$ yields the diffusion equation.
We refer to the review \cite{wei} for standard material and history related to applications of persistent random walks and the telegraph equation.  
Various versions or extensions of that telegraph equation \eqref{te} are naturally related to active motion as we will show; see also \cite{ind,cas}.  Similarly, studies in colored noise such as in \cite{fox,pha} may be viewed as predecessors of the research into active particles. A more recent version of noise representing activity is in \cite{far}.\\
Finally yet another relation may be found in studies of spin transport; see e.g. \cite{nel,ham,sun}.  There, as in the context of spintronics, spin-orbit coupling is the analogue of the active steering of translational motion through internal degrees of freedom.\\

  In the present paper, we take the internal degrees of freedom to be discrete, and we call them {\it spin} for short,  so that we imagine spatial point particles that can be in a finite number of internal states (spin values). 
To be more specific, we introduce the class of models in one dimension that we call {\it active}. We only deal here with independent particles. The state space is  $M=\mathbb{T} \times K$, where $\mathbb{T}\subset \mathbb{R}$ can stand for the circle $S^1$, an interval $[-\ell,\ell]$, the entire line or some discrete (lattice) version thereof. The set $K$ is finite and contains the $n$ possible (spin) values, In any event we call $x\in\mathbb{T}$ the (spatial) position and $\sigma\in K$ the spin.  The coupled dynamics (position--spin) in the form of an overdamped diffusion in one dimension is
\begin{eqnarray}\label{mode1}
\dot{x}_t - v(\sigma_t)= - \chi(\sigma_t)\, \frac{\id U}{\id x}(x_t,\sigma_t) + \sqrt{2D(\sigma_t)}\;\xi_t\\
\log\frac{k_x(\sigma,\sigma')}{k_x(\sigma',\sigma)} =  [U(x,\sigma) - U(x,\sigma') + F(x;\sigma,\sigma')]/T\nonumber\\
\sqrt{ k_x(\sigma,\sigma')\;k_x(\sigma',\sigma)} = \psi(x;\sigma,\sigma')\nonumber
\end{eqnarray}
where we need to add suitable boundary conditions on $x$ depending on the geometry.
The first line contains the driving velocity $v(\sigma)$ which depends on the spin $\sigma$.  A conservative coupling between position $x$ and spin $\sigma$ goes via the potential $U(x,\sigma)$.  The mobility coefficient is $\chi(\sigma)\geq 0$ and $T= D(\sigma)/\chi(\sigma) \geq 0$ (independent of $\sigma$) is the temperature of a thermal environment represented by standard white noise $\xi_t$. (Boltzmann's constant is set to one.)  The second and third line of \eqref{mode1} specify the transition rates for the spin.  The spin follows a Markov jump process with rate $k_x(\sigma,\sigma')$ for the transition $\sigma\rightarrow \sigma'$ at fixed (spatial) position $x$.  There is the antisymmetric $F(x;\sigma,\sigma')=-F(x;\sigma',\sigma)$ as a possible extra source of nonequilibrium driving in spin-space.  The $\psi(x;\sigma',\sigma)=\psi(x;\sigma,\sigma')$ is the symmetric activity part of the rates.  The coupling of translational motion and internal spin dynamics combined with the presence of the imposed velocity $v$ and/or of spin-driving $F$ is what makes the motion {\it active}.   Note also the difference with random walkers in a random environment; active particles carry the randomness related to possible bias or traps ``on their back.''\\

In the next section we start with active diffusion, both athermal and thermal, and we ask for the stationary density.  That analysis intersects with recently reported results in \cite{ind}.  We also discuss the reversibility or possible irreversibility in the spatial motion.  In particular we find an example where the spatial motion shows time--symmetry for all stationary two-time correlation functions, but not for higher--order time--correlations.  We continue with the violation of the Sutherland--Einstein relation even when the joint process is generalized reversible.  At the end of that Section we give an extension of the Arrhenius law to run-and-tumble particles.  The escape rate from a potential well over a barrier is computed and the dependence on the tumble frequency is manifested. In contrast with the case of pure diffusion the slope of the potential constitutes essential information for the exponential law.\\ 
In Section \ref{trr} we start by deleting the explicit velocity field $v(\sigma_t)$ in \eqref{mode1}; all nonequilibrium is then caused by the presence of the antisymmetric driving $F$ in spin-space. The natural question is to study the transmission of internal currents (for $\sigma_t$) to translational (directed) motion (for $x_t$), from which a velocity $v$ would emerge. It is interesting and useful at that moment to take into account a possible dependence of the mobility $\chi$ on spin $\sigma$.  In Section \ref{acti} the mobility $\chi(\sigma)\equiv 1$ in \eqref{mode1} is still constant independent of $\sigma$; in Section \ref{trr} we follow the idea that only at certain values of the spin it is possible for the walker to move forward or backward.  In other words, the spin values give kinetic constraints (no motion at $\sigma$ when $\chi(\sigma)=0$), which is in fact also similar to what may happen in molecular motors as movement there is only possible at certain chemo-mechanical configurations of the motor. We discuss there the issues of stalling and of the direction of the translational current, and how it depends on the symmetric activity parameters.\\
As will appear in the next section on active diffusion the telegraph equation \eqref{te} is naturally linked to active processes.  We give an extension, the thermal telegraph equation, in Section \ref{difc} (see equation \eqref{T-tel}).  We make a generalization to more than two lanes (general $n$ in \eqref{mode1}) in Section \ref{out}. \\
Active particles have by now been studied and reviewed from many perspectives and the discussion below has overlap with various other papers to which we refer.  The original contributions of the present paper concern (1) the discussion of (broken) time-reversal invariance (Section \ref{genrev}), (2) the violation of the Sutherland-Einstein relation as compatible with the fluctuation--dissipation relation for generalized reversible processes (Section \ref{sev}), (3) the derivation of a generalized telegraph equation for run-and-tumble processes with more than two driving velocities (Section \ref{out}), (4) the discussion on the Arrhenius law for run-and-tumble processes (Section \ref{ah}), and (5) the phenomenon of stalling for kinetically constrained active particles (Section \ref{trr}).

\section{Active diffusion}\label{acti}

To start we consider a single particle with position $x$ pushed with velocity field $c_\sigma(x)$ depending on spin $\sigma=\pm 1 $.  In other words, we take \eqref{mode1} with $\left[v(\sigma) - \chi(\sigma)\,U'(x,\sigma)\right] = c_x(\sigma)$ and $k_x(\sigma,-\sigma) = k_\sigma(x)$:
\begin{equation}\label{zerot}
\dot{x} = c_\sigma(x),\; \,\qquad \text{and jumps  } \sigma \longrightarrow -\sigma \text{ at rate } k_\sigma(x)
\end{equation}
where the $c_\sigma(x)$ are smooth in $x$ with $c_+\geq 0 \geq c_-,\, c_+ > c_-$, while the transition rates $k_\sigma(x) \geq \varepsilon >0$ are piecewise continuous in $x$.  The model is athermal for the moment, with $D=0=T$ in \eqref{mode1}.\\
The evolution of the joint probability density $(\rho_+(x,t),\,\rho_-(x,t))$  is
\begin{equation} \label{dyn}
\begin{cases}
& (\partial_t \rho_+)(x,t) = -\partial_x(c_+(x) \rho_+(x,t)) + k_-(x)\rho_-(x,t)-k_+(x)\rho_+(x,t)\\
& (\partial_t \rho_-)(x,t) = -\partial_x(c_-(x) \rho_-(x,t)) + k_+(x)\rho_+(x,t)-k_-(x)\rho_-(x,t)
\end{cases}
\end{equation}

The standard telegraph process \eqref{te} for the total density $\rho =\rho_+ + \rho_-$ corresponds to $c_\sigma = c\,\sigma$ for amplitude $c>0$ and activity parameter $k_\sigma= a =\eta/2$ and is then easily obtained from \eqref{dyn}. 

\subsection{Stationary density of athermal run-and-tumble processes}

Stationary distributions for independent active particles have been obtained in various cases; see e.g. \cite{mar,be}.
While completing this paper a very similar study appeared as \cite{ind}, dealing with run-and-tumble particles in one dimension.  In particular, a complementary treatment of the stationary density was given there at least in the thermal case.  We continue here first with the athermal case \eqref{zerot}, which was already considered in \cite{cv1,cv2}.\\

The stationary density  $(\rho_+,\rho_-)$ for \eqref{dyn} solves
\begin{equation} \label{stat}
\begin{cases}
& 0 = -(c_+ \rho_+)' + k_-\rho_--k_+\rho_+\\
& 0 = -(c_- \rho_-)' + k_+\rho_+-k_-\rho_-
\end{cases} 
\end{equation}
where boundary conditions ought to be added depending on the geometry.  For example these equations are impossible to verify on the circle unless $\oint k_-\,\rho_- = \oint k_+\,\rho_+$.
On the line, from adding the two equations in  \eqref{stat} and integrating over $x$,
\begin{equation} \label{res}
c_+\rho_++c_- \rho_- = \text{constant  } \,J
\end{equation}
where $J$ has the interpretation of being the total current in the direction of $x$.\\
Combining \eqref{res} with \eqref{stat} we get
\[
-(c_\sigma\rho_\sigma)'+\frac{k_{-\sigma}}{c_{-\sigma}} \left(J - c_\sigma \rho_\sigma\right)-k_\sigma \rho_\sigma=0
\]
with general solution 
\begin{equation} \label{expr1}
\rho_\sigma(x)= \frac{1}{c_\sigma(x)}e^{\phi(x)}\left[\frac{J}{2}+\sigma\,A+ J\,\int_0^x \frac{k_{-\sigma}(x')}{c_{\sigma}(x')c_{-\sigma}(x')}e^{-\phi(x)}\dx'\right]
\end{equation}
where $A$ is an integration constant and $\phi$ is defined through 
\begin{equation} \label{def}
\phi(0)=0,\qquad  \phi'=-\frac{k_+}{c_+}-\frac{k_-}{c_-}
\end{equation}
where boundary conditions will of course matter again for their (possible) solution. For the infinite line there may not be any stationary density.\\

For the dynamics on $[-\ell,\ell]$  we should let $c_+$ vanish at $x=\ell$ while $c_-$ should vanish at $x=-\ell$. Then, requiring that the densities $\rho_{\sigma}$ have a finite mass, the stationary density from \eqref{expr1} must have translational current $J=0$ and we get
\begin{equation} \label{sol2}
\rho_\sigma(x)=\sigma\,\frac{A}{c_\sigma(x)}e^{\phi(x)},\qquad A^{-1} =\int_{-\ell}^\ell\id x\, e^{\phi(x)}\,[\frac 1{c_+(x)}- \frac 1{c_-(x)}]
\end{equation}
Note that wherever $c_+(x)=-E c_-(x)>0$ for some constant $E>0$ (while the $k_{\sigma}$ remain arbitrary), the densities given by \eqref{sol2} are proportional to each other:
\begin{equation} \label{sol3}
\rho_+(x)=\frac{A}{c_+(x)}e^{\phi(x)} = \frac{1}{E}\rho_-(x) 
\end{equation}
We can for example take constant $k_\sigma=a>0$ and
\begin{equation}
c_{\sigma}(x)=\begin{cases}
0& \text{ when }x=\sigma\,\ell \\
\sigma\,c& \text{ otherwise}
\end{cases}
\end{equation}
Then, we find
\begin{eqnarray}\label{zer}
&&\rho_+(x)=\frac{1}{2+4\ell a/c}\left\{\delta(x-\ell)+\frac{a}{c}\right\}\qquad \rho_-(x)=\frac{1}{2+4\ell a/c}\left\{\delta(x+\ell)+\frac{a}{c}\right\}\\
&& \rho(x)=\rho_+(x)+\rho_-(x)=\frac{1}{2+4\ell a/c}\left\{\delta(x-\ell)+\delta(x+\ell)+\frac{2a}{c}\right\}
\end{eqnarray}
in agreement with equation (17) in \cite{ind}.\\

On the other hand, if we add a harmonic potential $V(x)=\kappa x^2/2$ by taking
\begin{equation} \label{case}
\begin{cases}
& c_+(x)=c-\kappa x=c-V'(x)\\
& c_-(x)=-c-\kappa x=-c-V'(x) \\
& k_+(x)=k_-(x)=a
\end{cases}
\end{equation}
we  must fix  $\kappa=c/\ell$ for $c_\sigma$ to vanish at $\sigma\ell$.  Integrating \eqref{def} and plugging the result into \eqref{sol2} yields
\begin{equation}\label{hatr}
\rho_\sigma(x)\propto \frac{\left(c^2-(\kappa x)^2\right)^{a/\kappa}}{c-\sigma\,\kappa x}
\end{equation}
Notice that the $\sigma=+1$ particles pile up around $\ell=c/\kappa$ while the $\sigma=-1$ particles pile up at the other side of the interval, as expected. That is an instance of breaking of Boltzmann statistics even for the position variable alone; see again under Section \ref{eqb}.

\subsection{To break or not to break time-reversal symmetry}\label{genrev}

Various questions can be asked about the time-reversibility of active processes. In particular, we can ask for the entropy production, cf. \cite{trev}, as will also appear at the end of Section \ref{sev}.  Yet, it is important to first specify precisely the question on reversibility, as we now indicate for the athermal processes above.\\

The simplest question is to ask whether the coupled Markov process \eqref{zerot} is reversible in its stationary distribution \eqref{sol2}.  That amounts to asking detailed balance for the coupled process, which is in general not true of course.  A formal but instructive proof (of that irreversibility) goes via the calculation of the generator $L^*$ of the time-reversed process.  Note that the backward generator $L$ of the Markov process \eqref{zerot} acts on a function $f_\sigma(x)$ as 
\[
(Lf)_\sigma(x)=c_\sigma(x) f_\sigma'(x) + k_\sigma(x)\left(f_{-\sigma}(x)-f_\sigma(x)\right)
\]
Consider now the forward generator
$L^{\dagger}$, the  adjoint of $L$.  For stationary distribution $\rho$ in \eqref{sol2} and any smooth function $f$, e.g. with $\sigma=+$,
\begin{eqnarray}
&& (L^{\dagger}(\rho f))_+ = -(c_+\rho_+f_+)' + k_-\rho_-f_- - k_+\rho_+f_+ \nonumber\\
&& = -c_+\rho_+f_+'- (c_+\rho_+)'f_++ k_-\rho_-f_- - k_+\rho_+f_+ \nonumber\\
&& = -c_+\rho_+f_+'- (k_-\rho_- - k_+\rho_+)f_++ k_-\rho_-f_- - k_+\rho_+f_+ \label{st1}\\
&& = -c_+\rho_+f_+'+ k_-\rho_-[f_--f_+] \nonumber\\
&& = \rho_+\left[-c_+f_+' - \frac{c_+k_-}{c_-}[f_--f_+]\right] \label{st2}
\end{eqnarray}
where \eqref{stat} was used in step \eqref{st1} and \eqref{sol2} in step \eqref{st2}.
From $L^{\dagger}$ we find the expression of the forward generator $L^*$ of the time-reversed Markov process via the formula $L^{\dagger}(\rho f) = \rho L^*f$ when we take $\rho$ to be the stationary density $\rho_\sigma(x)$ of \eqref{sol2}.  From \eqref{st2} we thus get
\begin{equation}\label{adj}
(L^*f)_\sigma=-c_\sigma f_\sigma' - \frac{c_\sigma k_{-\sigma}}{c_{-\sigma}}(f_{-\sigma}-f_\sigma)
\end{equation}
Therefore, $L\neq L^*$ in general which means that the stationary Markov process \eqref{zerot} is indeed not reversible.\\
On the other hand, the system does satisfy a {\it generalized} detailed balance if the velocities are antisymmetric, where we flip the spin as part of the kinematical time-reversal. Define the involution ${\cal I}(x,\sigma) = (x,-\sigma)$, and by extension $({\cal I}f)_\sigma(x) = f_{-\sigma}(x)$.  Then, 
\[
({\cal I} L^* {\cal I} f)_\sigma = -c_{-\sigma} f_\sigma' - \frac{c_{-\sigma} k_{\sigma}}{c_{\sigma}}(f_{-\sigma}-f_\sigma)
\]
so that ${\cal I} L^* {\cal I} = L$ when $-c_- = c_+$.  In that sense, when indeed $c_\sigma = E(x)\sigma$, then the joint position-spin process is time-reversal invariant in the same way a Hamiltonian dynamics is.\\  

Another aspect is to look at the stationary stochastic process of positions $X(t)$ (only) obtained by integrating out the spin $\sigma$. Then, the resulting or induced process is reversible, at least for the two-lane set-up treated so far.  That does not need to remain like that for $n>2$.\\
To illustrate that most easily, we take a model with $n=4$ (four lanes for spin $\sigma\in \{-2-1,1,2\}$) and with position on a (discrete) ring with $N$ sites.  The evolution is in discrete time given by
\begin{eqnarray}
&& X(t+1)=X(t)+\sigma(t)\mod N \nonumber\\
&&\sigma(t+1)=\begin{cases}
& \sigma(t) \text{ with probability }1-\epsilon\\
& G(\sigma(t)) \text{ with probability }\epsilon \text{ for }
\end{cases} \nonumber \\
&& G(-2)=1,\,G(1)=2,\,G(2)=-1,\,G(-1)=-2 \nonumber
\end{eqnarray}
Note that in the stationary process  all states $(X,\sigma)$ occur {\it a priori} with probability $\frac{1}{4N}$. One can also check that the probability of spin-history $\{(t,\sigma(t))\}_{0\leq t \leq {\cal T}}$ over time $[0,{\cal T}]$ is the same as the probability of $\{(t,-\sigma(t))\}_{0\leq t \leq {\cal T}}$. That implies that the probability to move forward a certain distance --- starting from $X_0$ -- in time ${\cal T}$ is the same as for moving backward the same distance, and in fact does not depend on that initial position $X_0$.  Therefore, all the two-time correlations in the stationary process associated to this evolution satisfy
\begin{eqnarray*}
&& \text{Prob}[X(t_0)=X_0,\,X(t_1)=X_1]=\underbrace{\text{Prob}[X(t_0)=X_0]}_{=\frac{1}{N}}\text{Prob}[X(t_1)=X_1|X(t_0)=X_0]\\
&&=\underbrace{\text{Prob}[X(t_0)=X_1]}_{=\frac{1}{N}}\text{Prob}[X(t_1)=X_0|X(t_0)=X_1]=\text{Prob}[X(t_0)=X_1,\,X(t_1)=X_0]
\end{eqnarray*}
Looking at two-time correlations therefore, the irreversibility of the original process goes unnoticed. At the level of three-point temporal correlations however, we have e.g.
\[\text{Prob}[X(t_0)=X_0,\,X(t_0+1)=X_0+1,\,X(t_0+2)=X_0+3]=\frac{1}{4N}\epsilon\]
while for the reversed process
\[\text{Prob}[X(t_0)=X_0+3,\,X(t_0+1)=X_0+1,\,X(t_0+2)=X_0]=0\]

That example can be mimicked in the continuous space and/or continuous time setting.  To the best of our knowledge it is the most simple example of a diffusive process not showing any time-reversal breaking on the level of two-time correlations and yet being irreversible. Remark that four is the minimal number of lanes where this higher-order irreversibility manifests itself; in a three-lane setting one seemingly again obtains a fully reversible process after integrating out the spin.

\subsection{Thermal run-and-tumble processes}\label{difc}
We now look at a finite temperature ($T$) version of \eqref{zerot}, which corresponds to the run-and-tumble processes of \cite{ind} with $\sigma=\pm 1$.  In other words we look now at \eqref{mode1} with $v(\sigma) - \chi(\sigma) U'(x,\sigma) = \sigma\,c$, $k_x(\sigma,-\sigma) = k_\sigma(x)$ and $D(\sigma) =T$:
\begin{equation}\label{nzerot}
\dot{x} = \sigma\,c + \sqrt{2T} \xi_t,\; \,\qquad \text{and jumps  } \sigma \longrightarrow -\sigma \text{ at rate } k_\sigma(x)\equiv a
\end{equation}
taking thus $c_+= c\geq 0, c_-=-c$.  Equation (16) in \cite{ind} gives the stationary density on a finite interval.  In the transient case we have the Smoluchowski equation for the spatial density $\rho$ which satisfies
\begin{equation} \label{T-tel}
(\partial_t -T\partial_x^2)^2\rho - c^2\partial_x^2\rho = -2a(\partial_t-T\partial_x^2 )\rho
\end{equation}
The derivation of \eqref{T-tel} goes as follows:\\
The Fokker-Planck equation for the probability density (we still write $\rho_{+}(x,t):=\rho(x,\sigma=+1,t)$ and $\rho_{-}(x,t):=\rho(x,\sigma=-1,t)$) is given by
\begin{equation}
\begin{cases}
& \partial_t \rho_+ = \partial_x \left(-c\rho_++T \partial_x \rho_+\right)+a(\rho_--\rho_+)\\
& \partial_t \rho_- = \partial_x \left(c\rho_-+T \partial_x \rho_-\right)+a(\rho_+-\rho_-)
\end{cases}
\end{equation}
That can be written equivalently as $L_+ \rho_+ = a(\rho_--\rho_+)=-L_-\rho_-$, where we defined the operators $L_{\pm}$ as 
\begin{equation}\label{begin}
L_{\pm}\nu = \partial_t \nu - \partial_x \left(\mp c\nu + T \partial_x \nu\right)
\end{equation}
One has
\begin{equation}
L_{-}L_{+}\nu = L_{+}L_{-}\nu = (\partial_t -T\partial_x^2)^2\nu - c^2\partial_x^2\nu
\end{equation}
So then
\begin{eqnarray*}
&& L_-L_+ \rho_+ = aL_-(\rho_--\rho_+) \\
&& =a^2(\rho_+-\rho_-)+aL_+\rho_+-2a(\partial_t-T\partial_x^2 )\rho_+\\
&& =-2a(\partial_t-T\partial_x^2 )\rho_+
\end{eqnarray*}
from which we conclude that
\begin{equation} \label{end}
(\partial_t -T\partial_x^2)^2\rho_+ - c^2\partial_x^2\rho_+ = -2a(\partial_t-T\partial_x^2 )\rho_+
\end{equation}
and one can verify that $\rho_-$ solves it as well. So the total density $\rho = \rho_++\rho_-$ solves \eqref{T-tel}.\\

There is another instructive way to write that thermal telegraph equation.  The equation \eqref{T-tel} is equivalent to the following system for functions $\pi(x,t), \rho(x,t)$,
\begin{equation}\label{hal}
\begin{cases}
& (\partial_t -T\partial_x^2)\rho = \pi \\
& (\partial_t -T\partial_x^2)\pi = c^2\partial_x^2\rho -2a\,\pi
\end{cases}
\end{equation}
Defining the energy functional
\begin{equation}
{\cal H}[\pi,\rho]:=\frac{1}{2}\int \id x\left(\pi^2+c^2(\partial_x\rho)^2\right)
\end{equation}
one can rewrite \eqref{hal} and thus also \eqref{T-tel} as an equation of motion for $(\rho,\pi)$ with a Hamiltonian plus a dissipative part,
\begin{equation}
\begin{cases}
& \partial_t\rho = \frac{\delta {\cal H}}{\delta \pi}-\frac{T}{c^2}\frac{\delta {\cal H}}{\delta \rho} \\
& \partial_t \pi = -\frac{\delta {\cal H}}{\delta \rho}-(2a-T\partial_x^2)\frac{\delta {\cal H}}{\delta \pi}
\end{cases}
\end{equation}
One now easily verifies for example that ${\cal H}$ is a Lyapunov functional.\\
One obtains the athermal telegraph equation of the previous subsection in the limit $T=0$.

\subsection{Equipartition breaking}\label{eqb}
We can try to capture the active particle in a harmonic potential
\[
V(x)=\frac{\kappa\, x^2}{2}
\]
Recall that a particle performing a detailed-balance dynamics would assume the density $\rho(x) \propto e^{-V(x)/T}$ and would therefore have its variance converging in time to $\langle x^2\rangle_\text{eq} =\frac{T}{\kappa}$. The environment temperature $T$ therefore is equal to $\kappa\,\langle x^2\rangle_\text{eq}$, an instance of the equipartition theorem.\\
For the active particle the stationary equation of motion now becomes
\begin{equation} \label{EOMstat}
\begin{cases}
& 0 = \partial_x \left(-[c-\kappa\,x]\rho_++T \partial_x \rho_+\right) + a\,(\rho_--\rho_+)\\
& 0 = \partial_x \left([c+\kappa\,x]\rho_-+T \partial_x \rho_-\right) + a\,(\rho_+-\rho_-)
\end{cases}
\end{equation}
One obtains three pairs of equations by multiplying \eqref{EOMstat} with $1, x$ and $x^2$ and integrating over $x\in \mathbb{R}$.
From those six equations, one finds
\begin{equation} \label{temp}
\kappa\,\langle x^2 \rangle = T + \frac{c^2}{\kappa+2a} =: T_\text{eff}(\kappa)
\end{equation}
which we can call an effective temperature, which is always greater than $T$ as expected.
For the first moments on the two lanes (so for fixed $\sigma=+1$ and $\sigma=-1$ respectively),
\begin{equation}
\langle x \rangle_+ = \frac{c}{\kappa+2a} = -\langle x\rangle_-
\end{equation}
which signals again that the stationary density in a symmetric trap acquires a bimodal character as a result of the activity, or shows non-Boltzmann statistics as mentioned  with formula \eqref{hatr}.  Non-Boltzmann distributions have of course been observed in active particles; cf. \cite{be,vo,ca}, and equivalence with other {\it effective} equilibrium potentials is discussed in \cite{far}.

\subsection{Violation of Sutherland-Einstein relation}\label{sev}
To obtain the diffusion constant $\cal D$ we release the process with initial data concentrated at $x=0$ and track $\langle x^2\rangle(t)$ for $t$ large. The relative probability of starting with $\sigma =1$ or $\sigma=-1$ is not important. 
Multiplying equation \eqref{T-tel} by $x^2$ and integrating, one gets
\[
\ddot{\langle x^2\rangle}-2c^2=-2a\dot{\langle x^2\rangle}+4aT
\]
For $t\to +\infty$, one then obtains $\langle x^2\rangle-(2T+\frac{c^2}{a})t \to$ Const. and the diffusion constant is
\begin{equation}\label{diffu}
{\cal D}:=\lim_{t \to \infty}\frac{\langle x^2\rangle(t)}{2t}
= T + \frac{c^2}{2a}
\end{equation}
reproducing Eq.~(12) in \cite{ind} which is obtained there from the exact expression for the time-dependent density $\rho(x,t) = \rho_+(x,t) + \rho_-(x,t)$.
The expression \eqref{diffu} implies that there is already diffusion at zero temperature $T=0$, making the situation reminiscent of the one in quantum mechanics; see e.g. \cite{qm1}.
Furthermore,  comparing with \eqref{temp}, \eqref{diffu} implies
\begin{equation}\label{ineq}
	T\leq T_\text{eff}(\kappa)\leq T_\text{eff}(0)={\cal D}
	\end{equation}
The effective temperature $T_\text{eff}(V)$ associated to an arbitrary confining potential $V$ probably always falls  between the same limits $T$ and ${\cal D}$. The inequalities \eqref{ineq} then hint that the validity of a Sutherland-Einstein relation may depend on the interpretation of temperature.  When taking the original temperature $T$ it will not be met for this system, and that is easy to verify now by calculating the mobility.\\

To see the mean velocity (or current) $\nu = \lim_{t \to \infty} \frac{\langle x\rangle(t)}{t}$ resulting from the application of an extra electric field $\cal E$, we modify the drift $c_{\sigma}=\sigma\,c$ to $c_{\sigma}= \sigma\,c + \cal E$.  A trivial calculation then yields that  $\nu = \cal E$.
The mobility defined as $\mu := \frac{\nu}{\cal E}$ is thus $1$.
Plugging in the calculated values, one obtains
\begin{equation} \label{Suth}
\frac{{\cal D}}{\mu T}=\frac{\cal D}{T} = 1 +\frac{c^2}{2a \,T} > 1
\end{equation}  
so that the Sutherland-Einstein relation is broken. 
Active particles are relatively more diffusive than their passive counterparts. \eqref{ineq} reveals that in \eqref{Suth} an ad-hoc replacement of $T$ by an effective temperature would not have changed that verdict. Paradoxically however there is a way to argue that the fluctuation-dissipation relation still has sway, as we show next.\\

There is an exact expression for the mobility at finite time $t$.  Suppose we add the field $\cal E$ at time zero, then to linear order in ${\cal E}\downarrow 0$,
\begin{equation}\label{vio}
\frac 1{{\cal E}}\,\langle x_t-x_0 \rangle^{\cal E} = \frac{1}{2T}\,\langle(x_t-x_0)^2\rangle^0 
- \frac{c}{2T}\,\int_0^t\id s\,\langle (x_t-x_0)\,\sigma_s\rangle^0
\end{equation}
That can be obtained via standard path-integration techniques; cf. \cite{sei}.  We see that the last term of \eqref{vio} is not zero and is responsible for the violation of the Sutherland-Einstein relation.  Yet the reference distribution (where ${\cal E}=0$) is generalized reversible (cf. Section \ref{genrev}) and for the involution ${\cal I}$ that was used there, the term 
\begin{equation}\label{ep}
\frac 1{T}\,\int_0^t\id s\,{\cal E} \,\,(\dot{x}_s - c\sigma_s)
\end{equation}
is anti-symmetric under time-reversal. The path-quantity \eqref{ep} is in fact nothing else than the path-dependent entropy flux during the time $[0,t]$ in the thermal environment at temperature $T$ due to the perturbation with ${\cal E}$.  Observe now that the linear response \eqref{vio} is given as a correlation between the observable $x_t-x_0$ and that entropy flux \eqref{ep}. Hence, while the Sutherland--Einstein relation is broken, at the same time we keep \eqref{vio} expressing the standard fluctuation--dissipation relation as valid close-to-equilibrium.  If, on the other hand, we interpret the spin as symmetric under kinematical time-reversal, then the last term in \eqref{vio} corresponds to the frenetic contribution in the linear response around steady nonequilibrium, and we find the breaking of the Sutherland--Einstein relation much as expected, \cite{sei}.\\
As a final remark we note that we can easily calculate the second term in \eqref{vio},
\[
\langle (x_t-x_0)\,\sigma_s\rangle^0 = \int_0^t\id u\,\langle \left(\sigma_u\,c + \sqrt{2T} \xi_u\right) \,\sigma_s\rangle^0 =  c\,\int_0^t\id u\,\langle \sigma_u\,\sigma_s\rangle^0
\]
which in the limit $t\uparrow \infty$ yields
\[
\frac 1{t} \int_0^t\id u \int_0^t\id s \,\langle \sigma_u\sigma_s\rangle^0  =\frac 1{t} \int_0^t\id u \int_0^t\id s \,e^{-2a|u-s|}\rightarrow \frac 1{a}
\]
Hence, the $t\rightarrow \infty$ limit of \eqref{vio} gives $\mu = {\cal D}/T - c^2/(2Ta)$ in accordance with \eqref{Suth}.

\subsection{Higher-order telegraph equation for the spatial density with multiple spin values}\label{out}
So far we have concentrated on run-and-tumble processes that have $n=2$ in \eqref{mode1}.  We can however take that dynamics also with multiple lanes, $n\geq2$.  It means that the driving velocities can take more than two values but we stick to the homogeneous case where they do not depend on position and where the potential $U=0$.\\

Suppose $\rho(x,\sigma;t)$ denotes the probability density for finding spin $\sigma$ and position $x $ at time $t$.  They obey a coupled system of Kolmogorov forward-equations,
\begin{equation}\label{3t}
(P\rho)(.,\sigma;.) := \sum_{\sigma'} P_{\sigma \sigma'}(\partial_t,\,\partial_x)\rho(.,\sigma';.)= 0
\end{equation}
where $P_{\sigma \sigma'}$ is a polynomial in derivatives to time $t$ and position $x$ and $P$ is the $n \times n$ matrix with the operator-entries $P_{\sigma \sigma'}$. The evolution equation \eqref{3t} is what you get for the densities of model \eqref{mode1} when $U=0$ and $k_x(\sigma,\sigma')$ does not depend on $x$: with $k_x(\sigma,\sigma') = a(\sigma,\sigma')$ and escape rate $\sum_{\sigma'} a(\sigma,\sigma') =\zeta(\sigma)$, we have
\[
P_{\sigma \sigma'}(G,H) = \delta_{\sigma,\sigma'}[G + v(\sigma) H - T H^2 + \zeta(\sigma)] - a(\sigma,\sigma')
\]
in \eqref{3t}.\\
The $P_{\sigma\sigma'}$ are linear maps on the space of smooth spatio-temporal densities and they mutually commute under map composition, so they generate a commutative algebra.  
Therefore the characteristic polynomial $q(\lambda) := \det (\lambda I - P)=c_n \lambda^n+\ldots+c_0$ is well defined.  When we plug the matrix $P$ itself into that polynomial, the result is the zero matrix (as stipulated by the Cayley-Hamilton theorem \cite{cay}): $q(P)=0$.
Hence, for all smooth functions $v(x,t,\sigma)$
\[0=q(P)v=c_n (P^nv)+\ldots+c_1 Pv+c_0 v \]
 But \eqref{3t} implies that $P\rho =0$. Therefore
\[0=q(P)\rho=[c_n P^{n-1}+\ldots+c_1]\circ \underbrace{P\rho}_{=0}+c_0 \rho = c_0 \rho=q(0)\rho=\det (-P)\rho=(-1)^n(\det P)\rho\]
So, for each $\sigma$,
\begin{equation}\label{3tt}
\left[(\det P)(\partial_t,\partial_x)\right]\rho(.,\sigma;.) = 0
\end{equation}
(provided one ascertains that densities solving \eqref{3t} remain smooth if given smooth initial data).   
By taking the sum, for the spatial density $\rho(x,t)= \sum_{\sigma} \rho(x,\sigma;t)$, we thus get
	\begin{equation}\label{aut}
	\left[(\det P)(\partial_t,\partial_x)\right]\rho = 0
	\end{equation}
If the diagonal entries $P_{\sigma \sigma'}$ in \eqref{3t} are first-order in $\partial_t$ (as is the case when the joint process is Markovian) while the off-diagonal entries are zeroth-order, then the $\partial_t$-order of equation \eqref{aut} is equal to $n$, the number of spin values.  The equation \eqref{aut} is the  generalization of the telegraph equation for general (finite) spin space.\\
For clarity let us illustrate it for $n=2$ and $a(\sigma,\sigma')= a\, (1-\delta_{\sigma,\sigma'})$ as in thermal run-and-tumble processes. Then,
\[
P_{\sigma \sigma'}(G,H) = \delta_{\sigma,\sigma'}[G + c \sigma H - T H^2 - a] + a(1-\delta_{\sigma,\sigma'})
\]
and det $P(G,H) = (G+cH-TH^2+a)(G=cH-TH^2+a) - a^2 = (G-TH^2)^2 + 2a(G-TH^2)- c^2H^2$. Putting $G=\partial_t, H= \partial_x$ we check that \eqref{aut} reproduces \eqref{T-tel}.\\
 Comparing to the derivation \eqref{begin}-\eqref{end}, the proof of \eqref{3tt} really amounts to an abstraction of the process of Gaussian elimination (e.g. eliminating $\rho_-$ to get a closed equation for $\rho_+$), with the only curiosity being that the relevant co\"efficients are mutually commuting linear operators instead of complex numbers.

\subsection{Arrhenius formula for run-and-tumble processes}\label{ah}

An interesting question is to estimate the time for active particles to traverse a potential barrier.  It is related to first-passage problems as considered also in \cite{ind}.  Here we look at the escape time from a local minimum in the potential.  The more general theory is known today as Kramers' escape rate theory but very few results are available for nonequilibrium systems.  Recall that in passive diffusive systems, without driving, the expected escape time $\tau_p$ is of the form
\begin{equation}\label{ve}
\tau_p \approx C \,\exp{\frac{\Delta}{T}},\qquad \text{in the regime where } \Delta/T \gg 1
\end{equation}
to overcome a potential-barrier of height $\Delta$ at temperature $T$, and $C$ is some pre-factor which may depend on more details of the potential shape and on temperature. In this paper we are only after the exponential form, the so-called Arrhenius behavior, ignoring the pre-factor. However in contrast with that case of pure/passive diffusion now and naturally the shape of the potential becomes important: the local slope must be compared with the strength of the velocity driving $c$. That is especially clear in the athermal case: if the slope of the potential exceeds $c$ on the right side of a potential well and dips below $-c$ on the left side, then at $T=0$ the particle remains localized inside the well whereas it will eventually escape when the slope remains below those thresholds. Obviously, in the passive athermal case $c=T=0$, the particle never escapes.\\

\begin{figure}[h]
	\centering
	\includegraphics[width=0.75\textwidth]{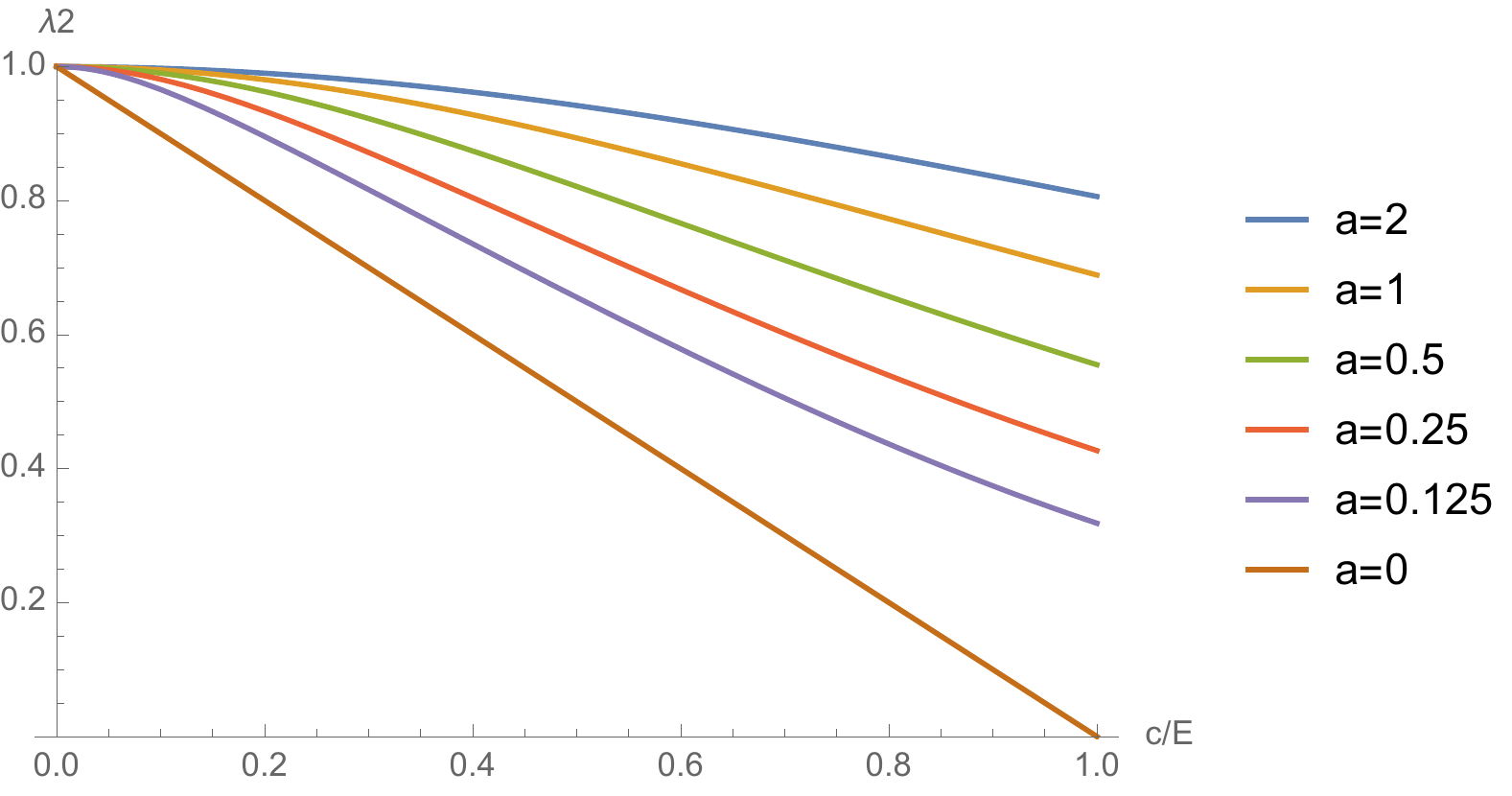}
	\caption{See formula \eqref{mr}--\eqref{la2}.  The plot gives $\lambda_2$ for ${\cal E}=T=1$ as function of the driving $c/{\cal E}$ for various amounts of persistence.  The escape time \eqref{mr} decreases with $c/{\cal E}$, and faster for larger persistence $\propto 1/a$. }\label{ar1}
\end{figure}

\begin{figure}[h]
	\centering
	\includegraphics[width=0.75\textwidth]{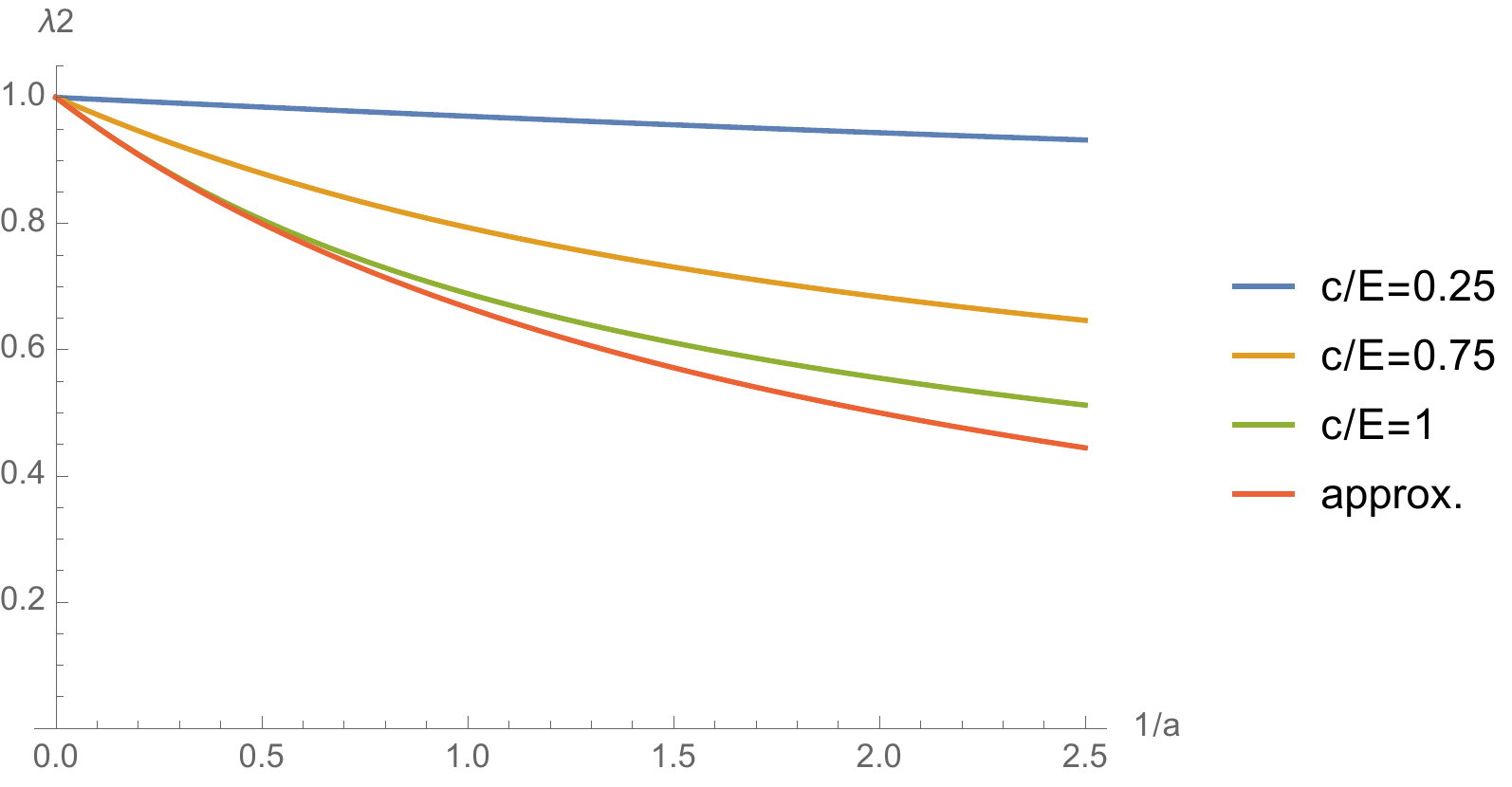}
	\caption{$\lambda_2$ as function of persistence $1/a$ for several driving amplitudes $c/{\cal E}$ while again ${\cal E}=T=1$.  Larger driving decreases the escape time \eqref{mr} for all values of persistence. The lower curve is $1/a \mapsto \frac{{\cal E}}{T+\frac{c^2}{2a}}$, in good agreement for small persistence, here compared with $\lambda_2$ for $c=1$.  For large persistence the asympotic values are $({\cal E}-c)/T$.}
	\label{ar2}
\end{figure}

To concentrate on the slope and for further simplicity we consider a linear potential $V(x) = {\cal E}\,|x|$ on the interval $[-h,h]$ where the potential height $V(h) = {\cal E}\, h$ will be taken very large while the slope ${\cal E}>0$ is fixed.  The dynamics for the position $x_t>0$ of the active particle is
\[
\dot{x}_t - c\sigma_t = -{\cal E} + \sqrt{2T}\,\xi_t
\]
while $\sigma_t$ is a Markov jump process on $\{+1,-1\}$ with rate $a_{\pm}$ for the transition $\pm 1 \rightarrow \mp 1$. There is a hard wall at the origin $x=0$ upon which the particle reflects without having a $\sigma$-transition (In the appendix it is clarified how this is precisely implemented).  At time zero $x_0=0$ with random initial spin $\sigma_0=\pm 1$ and we estimate the time $\tau$ required to reach $\pm h$ where the potential has height $\Delta := {\cal E} h$.  We 
 show in the Appendix \ref{arr} that, for fixed slope ${\cal E}>0$,
\begin{equation}\label{mr}
\tau(\Delta) = C\,\exp\left( {\lambda_2\,\frac{\Delta}{{\cal E}}}\right),\qquad \Delta\uparrow \infty
\end{equation}
with  $\log C$ sublinear in $\Delta$ and where, for the case $a_{\pm} = a$, 
\begin{equation}\label{la2}
\lambda_2= \frac{2{\cal E}}{3T} + 2\sqrt{p}\cos\left(\frac{1}{3}\left[\arccos\left(-q\, p^{-3/2}\right)-2\pi\right]\right)
\end{equation}
with
\[
\begin{cases}
& p=4{\cal E}^2/(9T^2)   - B/3\\
& q= -\frac{8}{27}{\cal E}^3/T^3 + \frac{1}{3} B\,{\cal E}/T + a \,\frac{{\cal E}}{T^2}\\
& B= ({\cal E}^2-c^2)/T^2 - 2a/T
\end{cases}
\]
We check immediately here that for $c=0=a$ (passive diffusion), we get $\lambda_2= {\cal E}/T$ letting \eqref{mr} to reproduce the Arrhenius law \eqref{ve}. The general qualitative behavior can be seen from Figs.~\ref{ar1} and \ref{ar2}.\\

The escape time is governed by the behavior of $\lambda_2$.  We can show that $\lambda_2$ decreases with temperature $T$ and increases with ${\cal E}/c$.  In fact $\lambda_2 \uparrow {\cal E}/T$ as $a\uparrow\infty$, and
$\lambda_2 \downarrow [{\cal E}- c]/T$ as $a\downarrow 0$ when ${\cal E} \geq c$.  In general $\lambda_2$ remains between those two limits.  They correspond to the values where the barrier height is either $\Delta_- :=({\cal E}-c)h>0$ or is $\Delta = {\cal E} h$ for passive diffusion (again, for $h$ large and ${\cal E}$ fixed):
\[
\tau_{T}^p(\Delta_-) \leq \tau(\Delta) \leq \tau_{T}^p\left(\Delta\right)
\]
with the superscript ``p'' refers to passive diffusion (at temperature $T$).  Especially the upper bound is not evident, as it says that activity helps the escape from a potential well and is also illustrated in greater details in Figs.~\ref{ar1} and \ref{ar2}.\\
If we take a passive diffusion with effective temperature $T_\text{eff}=T+\frac{c^2}{2a}$ as suggested by \eqref{diffu} and \eqref{Suth}, it always yields a smaller escape time than for the active diffusion at the physical temperature, but giving a good approximation for small persistence times ($a \to \infty$):
\begin{equation}\label{app}
\tau^p_{T_\text{eff}}(\Delta_-) \leq \tau(\Delta),\qquad  \lambda_2-\frac{\cal E}{T_\text{eff}} = O\left(\frac{{\cal E}c^4}{a^2T^3}\right)
\end{equation}
as is illustrated in the lowest curve in Fig.\ref{ar2}.\\

The result \eqref{mr} remains valid when $a_+\neq a_-$ with $\lambda_2$ being the middle (real) root of the polynomial
\begin{equation}
p(\lambda) = \lambda^3 - 2 \frac{{\cal E}}{T}\,\lambda^2+\left[\frac{{\cal E}^2-c^2}{T^2}-\frac{a_+}{T}-\frac{a_-}{T}\right]\lambda +
\frac{1}{T^2}\left[a_+({\cal E} + c) + a_-({\cal E} - c)\right]
\end{equation}
Formula \eqref{la2} gives that root in the case $a_-=a_+=a$.
If the transition rates comes to favour more the ``faster" lanes and penalizes more the slower ones, then the escape time $\tau$ gets smaller. On the other hand, if the slope gets ${\cal E}$ gets too small in the sense that
\[
0< \frac{{\cal E}}{c} \leq \frac{a_--a_+}{a_-+a_+}
\]
then the escape becomes ballistic (i.e $\tau \propto h$) and $\lambda_2\downarrow 0$.  That slope-dependence in the Arrhenius exponential is of course impossible for passive diffusion.

\section{Transmission of internal rotation}\label{trr}

\begin{figure}[h]
	\centering
	\includegraphics[width=.5\textwidth]{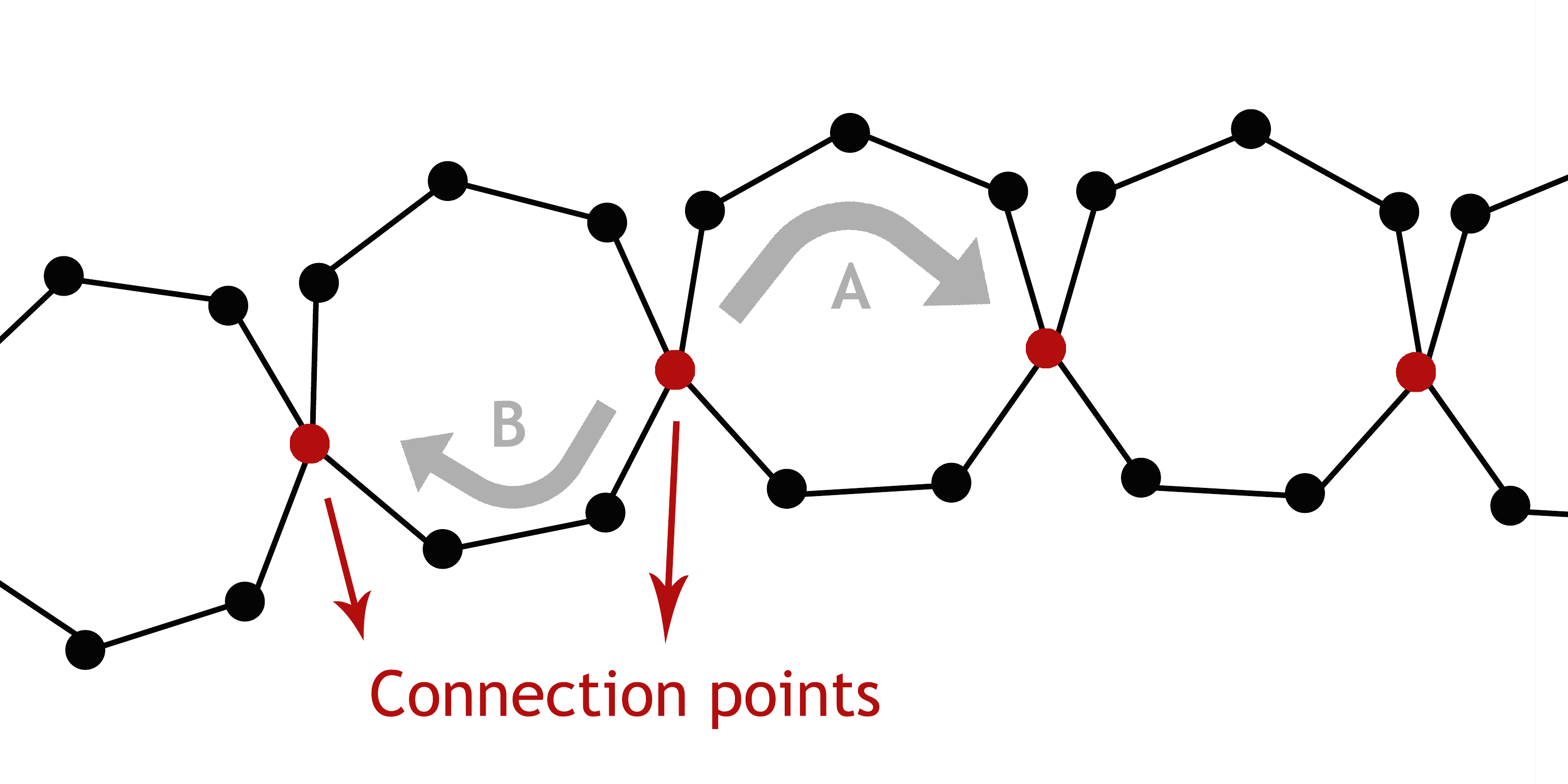}
	\caption{Internal states making the pearls on a necklace. Here 6 internal states are associated to each discrete position. Trajectory A expends as much entropy as trajectory B, leaving it undecided by the Second Law whether the spatial motion will be towards the right or towards the left.  Here 6 internal states are associated to each discrete position.  [Courtesy M. Stichelbaut, \cite{ms}.]}
	\label{figMC:intro}
\end{figure}

So far we have been following mostly the set-up of run-and-tumble processes.  The model \eqref{mode1} is however richer.  All the nonequilibrium driving can for example be delegated to the spin space. In that case we delete the driving velocity, putting $v(\sigma)=0$, and we hope it emerges effectively.  We are then dealing with model \eqref{mode1} taking the form
\begin{eqnarray}\label{modes}
\dot{x}_t =  - \chi(\sigma_t)\,\frac{\id U}{\id x}(x_t,\sigma_t) + \sqrt{2\chi(\sigma_t)\,T}\,\xi_t \\
\log\frac{k_x(\sigma,\sigma')}{k_x(\sigma',\sigma)} = \frac 1{T}\left[U(x,\sigma) - U(x,\sigma') + F(x;\sigma,\sigma')\right]\nonumber\\
\sqrt{ k_x(\sigma,\sigma')\;k_x(\sigma',\sigma)} = \psi(x;\sigma,\sigma')\nonumber
\end{eqnarray}
We repeat that the $F$ may break global detailed balance, by introducing an internal rotation in spin space.  In the simplest situation, for $n>2$, we may take $\sigma\in \mathbb{Z}_n$ (discrete ring) and put
\[
F(x;\sigma,\sigma\pm 1) = \pm \varepsilon, \qquad F(x;\sigma,\sigma') = 0 \text{ if }  \sigma'\neq \sigma\pm 1
\]
for nonequilibrium amplitude $\varepsilon>0$.  Similarly for the symmetric activity parameter we may put 
\[
\psi(x;\sigma,\sigma') = \phi_{\pm}(\sigma,\varepsilon)>0,\quad \text{ for } \sigma'=\sigma\pm 1,\qquad\text{ and zero otherwise}
\]
The first main question is then to understand if and how the internal rotation (on the spin ring $\mathbb{Z}_n$) gives rise to a spatial rotation for $x\in S^1$ on the circle. An easier case is obtained under the quasi-static limit where the spin-relaxation (for fixed $x$) to its (nonequilibrium) stationary distribution is much faster than the spatial motion.  For simplicity let us put still mobility $\chi(\sigma)=1$ constant and assume that $\phi_{\pm}(\sigma,\varepsilon)\gg 1$.  In the limit where the spin relaxes much faster than the position $x_t$, the equation of motion \eqref{modes} for the position becomes
\bea
\dot x_t = f(x_t) =  - \sum_{\sigma} \frac{\id U}{\id x} (x_t,\sigma)\; \rho(\sigma|x_t) + \sqrt{2\,T}\,\xi_t
\eea
for effective force $f(x)$, and $\rho(\sigma|x)$ is the stationary distribution on the spin $\sigma$ for a given position $x$ on the circle.  For $\varepsilon =0$ (equilibrium case) that effective force is derivable from a potential and the particle moves in a free energy landscape ${\cal F}(x) = -T \log \sum_{\sigma} \exp{-U(x,\sigma)}/T$. In general however (with $\varepsilon\neq 0$), the effective force is not derivable from a potential and has a rotational part $\oint f(x)\id x\neq 0$. In that case the steady spatial motion (on the circle) will show a current, transmitted from the rotation in spin-space.  The characteristics of that motion, e.g. direction and stability of fixed points, is strongly dependent on the choice of the activity parameter $\phi_\pm$ and how it varies with the driving $\varepsilon$.  We do not pursue that line here and some examples are given in \cite{up,nond}. \\
Instead we simplify the model mathematically to observe the effect beyond the quasi-static limit. At the same time we take serious the possible influence of spin-dependent mobility $\chi(\sigma)$. 
 The point is that spatial motion often (in the so called power stroke for molecular motors) is possible at all only from certain internal states, i.e., $\chi(\sigma) = 0$ for most spin states $\sigma$.
 To make the point more clearly we put the position $x_t$ also on a discrete ring and we study motion on structures which resemble necklaces as in Fig.\ref{figMC:intro}, \cite{ms}. There we see heptagons of internal spin states, $\sigma \in \mathbb{Z}_6$. 
To be specific and referring to Fig.\ref{figMC:intro} for the general situation, we think of the connection points being the possible spatial positions and there are exactly three internal states (out of the 6 in Fig.\ref{figMC:intro}) allowing movement to another spatial position.  Let us now fix the Markov jump process by giving the transition rates over each bond in Fig.\ref{figMC:intro}.  Supposing that there are $\ell$ bonds in trajectory A and $m$ bonds in trajectory B, we put for all bonds in trajectory A, the jump rate $k(\sigma,\sigma\pm 1) = \exp(\pm m\varepsilon)$ and for all bonds in trajectory B, $k(\sigma,\sigma\pm 1) = \varphi\,\exp(\pm \ell\varepsilon)$.
Note that the entropy flux for a journey along trajectory A is equal to the one associated to trajectory B:
\begin{equation}
{ k(x,0,1) k(x,1,2) \cdots k(x,\ell-1,\ell)   \over  k(x,\ell,\ell-1) \cdots k(x,2,1) k(x,1,0) } = \exp \varepsilon 
\end{equation}
for the trajectory A and a similar calculation for the trajectory B.\\

At every vertex, any outflow through some bond is exactly compensated by an inflow through a bond on the opposite side. As a consequence, the stationary density $\rho$ is uniform.
The physical current (say to the right in Fig.\ref{figMC:intro}) is then given by
\begin{eqnarray}
J_{\rightarrow}(\varphi,\varepsilon) &=& \rho\left\{\exp(m\varepsilon)-\exp(-m\varepsilon)-\varphi [\exp(\ell\varepsilon)-\exp(-\ell\varepsilon)]\right\}\nonumber\\
&=& 2\rho\left\{\sinh(m\varepsilon)-\varphi \sinh(\ell\varepsilon)\right\}
\end{eqnarray}
It is clear from that expression that
\begin{enumerate}
	\item When $\ell=m$, (symmetric pearls), $J_{\rightarrow}$ is only zero when $\varepsilon=0$ (equilibrium) or $\varphi=1,\,\varepsilon$ arbitrary. 
	\item When $\ell>m$, the current vanishes when either $\epsilon=0$ or otherwise when $\varphi=\frac{\sinh(n\varepsilon)}{\sinh(m\varepsilon)}$. Inverting the latter relation, we see that for all $\varphi$ strictly between $0$ and $1$, there exists a $\varepsilon_{crit}(\varphi)$ such that $J_{\rightarrow}(\varphi,\varepsilon_{crit}(\varphi))=0$ (stalling). 
\end{enumerate}
  We thus see that for fixed driving $\varepsilon$ the direction of the spatial current reverses as a function of the symmetric activity rate $\varphi$.
A similar dependence on the activity parameter has been shown in the model 3C in \cite{mmo}.


\section{Conclusions}
Active particles can be modelled usefully as diffusions or random walkers coupled to internal {\it spin} degrees of freedom.  For run-and-tumble processes a form of generalized reversibility holds for the total system.  Yet, the Sutherland--Einstein relation between diffusion constant and mobility is broken.  In the case or run-and-tumble processes with multiple spin values and corresponding driving velocities we obtained a generalized telegraph equation for the evolution of the spatial density. We also studied the modification of the Arrhenius formula where activity is in general seen to enhance the escape from a potential well.  We pointed out the importance of the (local) slope of the potential as a typical nonequilibrium effect.  Finally, we studied models where the coupling between spin and position becomes kinetically constrained as in molecular motors. There the asymmetry in activity parameters in the spin dynamics may govern the direction of the spatial current.  \\

\noindent{\bf Acknowledgment: We thank Urna Basu for useful discussions.}

\newpage
\appendix
\section{Arrhenius formula for run-and-tumble particles} \label{arr}

We collect here details about the derivation of \eqref{mr}.\\
 
When we have a run-and-tumble process on the state space $\mathbb{R}\times\{1,-1\}$ with driving velocity $c(x,\sigma)$ and spin-transition rates $a(x,\sigma)$ the backward generator $L$ is of the form
\begin{equation} \label{process}
(Lf)(x,\sigma)=T(\partial_x^2f)(x,\sigma)+c(x,\sigma)(\partial_xf)(x,\sigma)+a(x,\sigma)[f(x,-\sigma)-f(x,\sigma)]
\end{equation}
The expectation value $\tau(x,\sigma)$ of the time-span required to hit $(-\infty,-h]\cup [h,+\infty)$ if the process \eqref{process} is initialized at $(x,\sigma)$ ($x \in (-h,h)$) must obey the boundary value problem
\begin{equation}\label{BVP1}
\begin{cases}
& L\tau=-1 \\
& \lim_{x \to h}\tau(x,\pm 1)=0=\lim_{x \to -h}\tau(x,\pm 1)
\end{cases}
\end{equation}
If  $c$ transforms antisymmetric around $x=0$ the solution to \eqref{BVP1} is symmetric under the same reflection. That implies that \eqref{BVP1} yields the same solution on $(0,h)$ as 
\begin{equation}\label{BVP2}
\begin{cases}
& L\tau=-1 \\
& \lim_{x \to 0+}(\partial_x\tau)(x,\pm 1)=0=\lim_{x \to h-}\tau(x,\pm 1)
\end{cases}
\end{equation}
This is precisely what we do in the derivation of \eqref{mr}. The hard-wall condition is simply implemented by defining the process $(x_t,\sigma_t)$ considered there as $(|y_t|,\sigma_t)$ where
\begin{equation}
\dot{y}_t-c\sigma_t\theta(y_t)=-{\cal E}\theta(y_t)+\sqrt{2T}\xi_t
\end{equation}
where $\theta$ is the standard Heaviside-function. The latter process has the necessary antisymmetry to have \eqref{BVP1} reduce to \eqref{BVP2}.
Abbreviating $\tau(.,\pm1)=:\tau_{\pm}$, $a(.,\pm 1)=:a_{\pm}$, our problem now consists of solving the following system of ODE's (to be solved on the interval $(0,h)$):
\begin{equation}\label{sys3}
\begin{cases}
& T\tau_+''+(c-{\cal E})\tau_+'+a_+[\tau_--\tau_+]=-1\\
& T\tau_-''+(-c-{\cal E})\tau_-'+ a_-[\tau_+-\tau_-]=-1\\
& \tau_+'(0)=\tau_-'(0)=\tau_+(h)=\tau_-(h)=0
\end{cases}
\end{equation}
The solution is in general of the form
\begin{equation}\label{general}
\begin{cases}
& \tau_+(x)=\tau_{+,\text{part}}(x) + \mu_{0}+\sum_{j=1}^3\mu_{j}e^{\lambda_j x}\\
& \tau_-(x)=\tau_{-,\text{part}}(x) + \mu_{0}+\sum_{j=1}^3\mu_{j}z_je^{\lambda_j x}
\end{cases}
\end{equation}
wherein
\begin{enumerate}
	\item The complex numbers $\{\lambda_j\}_{1\leq j \leq 3 }$ are the roots of the polynomial
	\begin{eqnarray}\label{pol}
	&& p(\lambda):=\frac{1}{\lambda}
	\begin{vmatrix}
	T\lambda^2+[c-{\cal E}]\lambda - a_+ & a_+ \\
	a_- & T\lambda^2 -[c+{\cal E}]\lambda - a_-
	\end{vmatrix} \\ 
	&&=T^2\left\{\lambda^3 - \frac{2{\cal E}}{T} \lambda^2+\left[-\frac{c^2-{\cal E}^2}{T^2}-\frac{a_+}{T}-\frac{a_-}{T}\right]\lambda -\left[-\frac{a_+}{T}\frac{c+{\cal E}}{T}+\frac{a_-}{T}\frac{c-{\cal E}}{T}\right]\right\} \nonumber \\
	&&=T^2\left\{\lambda^3 - \frac{2{\cal E}}{T}\lambda^2+\left[\frac{{\cal E}^2-c^2}{T^2}-a_+/T-a_-/T\right]\lambda +\left[a_+/T\,\frac{{\cal E} + c}{T} + a_-/T \,\frac{{\cal E} - c}{T}\right]\right\}\nonumber
	\end{eqnarray}
	The pairs $\{(1,z_{j})\}_j=\{(1,\frac{a-(c-{\cal E})\lambda_j-T\lambda_j^2}{a})\}_j$ are the associated generalized eigenvectors.
	\item $(\mu_j)_{0\leq j\leq 4}$ are real parameters which are determined by the boundary conditions (note: 4 boundary conditions and 4 parameters)
	\item $\tau_{\sigma,\text{part}}$ are a certain ``particular'' solution that absorb the $-1$'s in the right-hand side of the system  \eqref{sys3}. Let us agree on the following choice:
	\begin{equation}\label{part}
	\begin{cases}
	& \tau_{+,\text{part}}(x)=A+Dx 
	=\frac{-2c}{-a_+(c+{\cal E}) + a_-(c-{\cal E})}-\frac{a_+ + a_-}{-a_+(c+{\cal E}) + a_- (c-{\cal E})} x
	 =A+ \frac{a_+ + a_-}{a_+ ({\cal E} + c) + a_-({\cal E} -c)}x\\
	& \tau_{-,\text{part}}(x)=Dx
	\end{cases}
	\end{equation}
\end{enumerate}
To proceed, it is worthwhile to observe that the characteristic polynomial \eqref{pol} must have three different real roots 
\begin{equation} \label{sort}
\lambda_1> ({\cal E} + c)/T \geq \lambda_2 \geq ({\cal E}-c)/T >\lambda_3, 
\end{equation}
Indeed, for $a_{\pm}\,c>0$ (active diffusion),
\begin{equation}
\begin{cases}
& p(\frac{{\cal E}-c}{T})=2\,a_+\,c > 0 \\
& p(\frac{{\cal E}+c}{T})= -2\,a_-\,c<0
\end{cases}
\end{equation}
so that the intermediate value and the large-$\lambda$ asymptotics of a cubic function yield the statement. In the marginal case $c=0$ (passive diffusion), the roots are given by $\{\xi,\frac{\xi}{2}\pm\sqrt{\left(\frac{\xi}{2}\right)^2+a_+/T + a_-/T}\}$ with $\xi={\cal E}/T$, which are three different real numbers, the middle one again being $\lambda_2={\cal E}/T$.
\\\\
The boundary conditions of \eqref{sys3}, written in terms of the expressions \eqref{general} \eqref{part}, take the form
\begin{equation}\label{BC2}
\begin{cases}
& 0=D+\sum_{j=1}^3\lambda_j\mu_j \\
& 0=D+\sum_{j=1}^3\lambda_jz_j\mu_j\\
& 0=A+Dh+\mu_0+\sum_{j=1}^3e^{\lambda_jh}\mu_j\\ & 0=Dh+\mu_0+\sum_{j=1}^3e^{\lambda_jh}z_j\mu_j
\end{cases}
\end{equation}
By Cramer's rule, the solution for $\mu_0$ is given by
\begin{eqnarray}
&& \mu_0 = \frac{\begin{vmatrix}
	-D& \lambda_1 & \lambda_2 & \lambda_3 \\
	-D& \lambda_1z_1 & \lambda_2z_2 & \lambda_3z_3 \\
	-A-Dh& e^{\lambda_1h} & e^{\lambda_2h}& e^{\lambda_3h} \\
	-Dh & e^{\lambda_1h}z_1 & e^{\lambda_2h} z_2& e^{\lambda_3h} z_3
	\end{vmatrix}}{\begin{vmatrix}
	0& \lambda_1 & \lambda_2 & \lambda_3 \\
	0& \lambda_1z_1 & \lambda_2z_2 & \lambda_3z_3 \\
	1& e^{\lambda_1h} & e^{\lambda_2h}& e^{\lambda_3h} \\
	1& e^{\lambda_1h}z_1 & e^{\lambda_2h} z_2& e^{\lambda_3h} z_3
	\end{vmatrix}} \approx -D\frac{(z_3-1)\lambda_3(z_2-z_1)e^{\lambda_1 h}e^{\lambda_2 h}}{(z_3-z_2)\lambda_2 \lambda_3(z_1-1)e^{\lambda_1 h}} \nonumber\\
&& =\frac{D(1-z_3)(z_2-z_1)e^{\lambda_2 h}}{(z_3-z_2)(z_1-1)\lambda_2} \label{mu0}
\end{eqnarray}
where the approximation becomes better provided $\lambda_2>0$ (In the special case where $\lambda_2 \leq 0$ one can calculate that $\tau \propto h$, i.e. the escape acquires a ``ballistic" speed) and with increasing $h$, in the sense that the inequality $\lambda_1 >\lambda_2 >\lambda_3$ becomes amplified to
\begin{equation} \label{sort2}
(\lambda_1-\lambda_2)h \gg 1  \qquad (\lambda_2-\lambda_3)h \gg 1.
\end{equation}
A similar computation reveals that $\mu_{1,2,3}$ are only of order $\max\{e^{\lambda_3 h}, e^{(\lambda_2+\lambda_3-\lambda_1) h}\}$ and due to \eqref{sort2}, in the large-$h$ limit this is much smaller than \eqref{mu0}. So in this large-$h$ limit $\tau_{\pm}(x=0)\approx \mu_0$. 
The careful reader will want to ensure that the prefactor in \eqref{mu0} has a nonzero numerator $(1-z_3)(z_2-z_1)$: $1-z_3=-a^{-1}(\lambda_3^2+(c-{\cal E})\lambda_3)>0$ since $\lambda_3<0$. Likewise, one can verify that in general $z_2-z_1>0$. Wrapping everything together,
\begin{equation}\label{fin}
	\begin{cases}
	& \tau_{\pm}\approx \frac{D(1-z(\lambda_3))(z(\lambda_2)-z(\lambda_1))e^{\lambda_2 h}}{(z(\lambda_3)-z(\lambda_2))(z(\lambda_1)-1)\lambda_2}\\
	& z(\lambda):= -\frac{T\lambda^2+(c-{\cal E})\lambda - a_+}{a_+}=-\frac{\lambda^2 T- ({\cal E}-c)\lambda - a_+}{a_+} \\
	& D:=-\frac{a_+ + a_-}{\alpha_+c_- + a_- c_+}= \frac{a_+ + a_-}{a_+ ({\cal E}+c) + a_-({\cal E}-c)} \\
	& \lambda_1 >\lambda_2 >\lambda_3 \text{ roots of the polynomial \eqref{pol}}
	\end{cases}
\end{equation}


\begin{thebibliography}{10}

\bibitem{sei}
M.~Baiesi, C.~Maes and B.~Wynants, The modified Sutherland-Einstein relation for diffusive non-equilibria. Proc. Royal Soc. A {\bf 467}, 2792--2809 (2011).

\bibitem{up}
U.~Basu, C.~Maes and K.~{Neto\v{c}n\'{y}}, How Statistical Forces Depend on the Thermodynamics and Kinetics of Driven Media. Phys. Rev. Lett. {\bf 114}, 250601 (2015).

\bibitem{be}
C.~Bechinger, R.~Di Leonardo, H.~L\"owen, C.~Reichhardt, G.~Volpe and  G.~Volpe, Active particles in complex and crowded environments. Reviews  of Modern Physics
{\bf 88}, 045006 (2016).

\bibitem{cv2}
I.~Bena, C.~Van den Broeck, R.~Kawai and K.~Lindenberg, Nonlinear Response With Dichotomous Noise. Phys. Rev. E {\bf 66}, 045603(R) (2002).

\bibitem{cv1}
I. Bena, C. Van den Broeck, R.~Kawai and K.~ Lindenberg,
Drift by Dichotomous Markov Noise. Phys. Rev. E {\bf 68}, 041111 (2003).

\bibitem{cas}
P.~Castro--Villarreal and F.~J.~Sevilla, Active motion on curved surfaces.
{\tt arXiv:1712.04619 [cond-mat.stat-mech]}	

\bibitem{ca}
M.E.~Cates and J.~Tailleur, When are active Brownian particles and run-and-tumble particles equivalent? Consequences for motility-induced phase separation.
Europhysics Letters	{\bf 101}, 20010 (2013).

\bibitem{pdb}
P.~de Buyl, A.S.~Mikhailov and R.~Kapral, Self-propulsion through symmetry breaking.
EPL (Europhysics Letters) {\bf 103}, 60009 (2013).
	
\bibitem{far}
\'E.~Fodor, C.~Nardini, M.E.~Cates, J.~Tailleur, P.~Visco and F.~van Wijland, How far from equilibrium is active matter? Phys. Rev. Lett. {\bf 117}, 038103 (2016).

\bibitem{mar}
 \'E.~Fodor, M. Cristina Marchetti, The statistical physics of active matter: from self-catalytic colloids
  to living cells. Lecture notes for the international summer school ``Fundamental
  Problems in Statistical Physics'' 2017 in Bruneck.  arXiv:1708.08652v3 [cond-mat.soft] 

\bibitem{fox}
R.F.~Fox, Uniform convergence to an effective Fokker-Planck equation for weakly colored noise.
Phys. Rev. A {\bf 34},	4525 (1986).

\bibitem{cay}
P.~Garrett, \textit{Abstract algebra},  NY: Chapman and Hall/CRC, 2007.

\bibitem{gold}
S.~Goldstein, On diffusion by discontinuous movements, and on the telegraph equation.  Q.J. Mech. Appl. Math. IV, 129--156 (1951).

\bibitem{pha}
P.~H\"anggi and P.~Jung, Colored Noise in Dynamical Systems.
Advances in chemical physics
{\bf 89}, 239--326 (1995).

\bibitem{ham}
N.J.~Harmon and M.E.~Flatt\`e, Spin relaxation in materials lacking coherent charge transport. Phys. Rev. B {\bf 90}, 115203 (2014).

\bibitem{Kac}
M.~Kac, A stochastic model related to the Telegrapher's Equation. Rocky Mountain J. Math. {\bf 4}, 497 (1974).

\bibitem{mmo}
C.~Maes, What decides the direction of a current?  Mathematics and Mechanics of Complex Systems {\bf 3-4}, 275--295 (2016).	

\bibitem{nond}
C.~Maes, Non-Dissipative Effects in Nonequilibrium Systems. SpringerBriefs in Complexity,  ISBN 978-3-319-67780-4 (2018).

\bibitem{ind}
Kanaya Malakar, V. Jemseena, Anupam Kundu, K. Vijay Kumar, Sanjib Sabhapandit, Satya N. Majumdar, S. Redner, Abhishek Dhar, 
Steady state, relaxation and first-passage properties of a run-and-tumble particle in one-dimension.
{\tt  	arXiv:1711.08474 [cond-mat.stat-mech]}

\bibitem{trev}
C.~Nardini, \'E.~Fodor, E.~Tjhung, F.~van Wijland, J.~Tailleur, M.~E.~Cates, Entropy production in field theories without time reversal symmetry: Quantifying the non-equilibrium character of active matter. Phys. Rev. X {\bf 7}, 021007 (2017). 

\bibitem{qm1}
G.~N.~Ord, The Schr\"odinger and Dirac Free Particle Equations without Quantum Mechanics. Annals of Physics {\bf 250}, 51--62 (1996).

\bibitem{vo}  
P.~Romanczuk, M.~B\"ar, W.~Ebeling, B.~Lindner and L.~Schimansky-Geier, Active brownian particles. The European Physical Journal Special Topics {\bf 202}, 1--162 (2012).
	
\bibitem{nel}
S.~Schott, E.R.~McNellis, C.B.~Nielsen, H.-Y.~Chen, S.~Watanabe, H.~Tanaka, I.~McCulloch, K.~Takimiya, J.~Sinova and H.~Sirringhaus, Tuning the effective spin-orbit coupling in molecular semiconductors. Nature Communications {\bf 8}, 15200 (2017).
	
\bibitem{ms}
M.~Stichelbaut, Steady currents in necklaces and ratchet systems. Master thesis KU Leuven 2016-2017.

\bibitem{sun}
D.~Sun ali Sun, Kipp J. van Schooten, M.~Kavand, H.~Malissa, Chuang Zhang, M.~Groesbeck, C.~Boehme and Z.~Valy Vardenyet, Inverse spin Hall effect from pulsed spin current in organic semiconductors with tunable spin–orbit coupling. Nat. Mater. {\bf 15}, 863--869 (2016).

\bibitem{wei}
G.H.~Weiss, Some applications of persistent random walks and the telegrapher's equation. Physica A {\bf 311}, 381--410 (2002).



	
		
	
\end{thebibliography}
\end{document}